\documentclass[review]{elsarticle}

\usepackage{hyperref,graphicx,booktabs,multirow,tabularx,array}

\begin{document}

\begin{frontmatter}

\title{Same Influenza, Different Responses: Social Media Can Sense a Regional Spectrum of Symptoms}

\author[mymainaddress,mysecondaryaddress]{Siqing Shan}
\author[mymainaddress,mysecondaryaddress]{Yingwei Jia}
\author[mymainaddress,mythirdaddress]{Jichang Zhao\corref{mycorrespondingauthor}}
\cortext[mycorrespondingauthor]{Corresponding author}
\ead{jichang@buaa.edu.cn}

\address[mymainaddress]{School of Economics and Management, Beihang University, Beijing, China}
\address[mysecondaryaddress]{Beijing Key Laboratory of Emergency Support Simulation Technologies for City Operation, Beijing, China}
\address[mythirdaddress]{Beijing Advanced Innovation Center for Big Data and Brain Computing, Beijing, China}

\begin{abstract}
Influenza is an acute respiratory infection caused by a virus. It is highly contagious and rapidly mutative. However, its epidemiological characteristics are conventionally collected in terms of outpatient records. In fact, the subjective bias of the doctor emphasizes exterior signs, and the necessity of face-to-face inquiry results in an inaccurate and time-consuming manner of data collection and aggregation. Accordingly, the inferred spectrum of syndromes can be incomplete and lagged. With a massive number of users being sensors, online social media can indeed provide an alternative approach. Voluntary reports in Twitter and its variants can deliver not only exterior signs but also interior feelings such as emotions. These sophisticated signals can further be efficiently collected and aggregated in a real-time manner, and a comprehensive spectrum of syndromes could thus be inferred. Taking Weibo as an example, it is confirmed that a regional spectrum of symptoms can be credibly sensed. Aside from the differences in symptoms and treatment incentives between northern and southern China, it is also surprising that patients in the south are more optimistic, while those in the north demonstrate more intense emotions. The differences sensed from Weibo can even help improve the performance of regressions in monitoring influenza. Our results suggest that self-reports from social media can be profound supplements to the existing clinic-based systems for influenza surveillance.
\end{abstract}

\begin{keyword}
Influenza monitoring \sep social media \sep spectrum of symptoms \sep regional differences \sep regression models
\end{keyword}

\end{frontmatter}


\section{Introduction}

Influenza is a highly contagious and rapidly spreading acute respiratory infection caused by the influenza virus. It is common but sometimes may become a severe or fatal disease in specific areas due to ubiquity of urbanization trends (\cite{Florian}\cite{Cameron}). Studies have shown that the seasonal and regional differences in influenza surveillance results are of great significance for understanding the epidemic details of influenza and the selection of vaccinations (\cite{Nguyen}\cite{Beaut}). Nevertheless, the traditional method for influenza surveillance is collecting information through various sentinel hospitals, by which it is not possible to obtain complete and effective information from patients in time to help control the epidemic (\cite{Rolfes}\cite{Bednarska}). In the meantime, the subjectivity of face-to-face inquiries might result in a biased spectra of symptoms, while too much emphasis on exterior signs may also overlook the interior suffering of the patients.

To compensate for the delay of traditional means for disease surveillance, a search engine was first excavated from the Internet to help monitor influenza trends (\cite{Ginsberg}\cite{Lazer}). Shortly after the boom of social media, Twitter and its variants became new and credible data sources in disease surveillance. Aside from their rich demographics, such as geography, gender, age and occupation, users of social media are also deeply embedded in social networks that are composed of following links (\cite{Abbasi}) or interaction connections. In the manner of self-reporting and word-of-mouth, content in the form of texts, photos and URLs are shared and disseminated, thereby accumulating digital footprints that can be unprecedented windows into understanding human behaviors. In particular, for the study of influenza, social networks offer the opportunity to trace the epidemic paths, and the sophisticated signals conveyed within content, not only factual descriptions but also mental statuses, make the comprehensive profiling of symptoms possible. According to the review by Charles-Smith (\cite{Charles}), Twitter was selected as the primary data source for 81\% of influenza-related explorations, and with the help of social media, 45\% of the previous studies of disease surveillance have tried to predict the spread of influenza (\cite{Moritz}\cite{Fred}\cite{Marque}). However, whether social media can sense a regional spectrum of symptoms, especially from views including both exterior and interior signs, is rarely examined.

Indeed, compared with other surveillance approaches, for example, collecting patients’ information from outpatient services, social media offers natural, voluntary "reports" to identify epidemiological symptoms and gain a spectrum of rich dimensions, including geography, social networks, and individual attributes. In particular, the conventional outpatient data do not reflect the psychological responses of the patients to the disease, while these internal reactions are in fact critical to understanding patients’ statuses and developing treatment plans. Landmark studies have demonstrated the critical contributions of taking patients’ emotions into account for disease treatment (\cite{Veruska}\cite{Askelund}\cite{Mellblom}). More inspiring, it is also found that for treating influenza, approaches considering moods might vary across populations, for example, in the military and controlled populations (\cite{Volkova}). This finding implies that for different regions, the differences beyond the same influenza would inherently help present exclusive treatments. Nevertheless, to the best of our knowledge, the exact picture of the regional differences in response to influenza is still missing.

With its vast territory and complex climatic conditions, China faces an extremely serious situation regarding influenza prevention and control. Specifically, the seasonality of influenza activities and the variations in influenza symptoms across regions greatly challenges the existing systems (\cite{Yu}\cite{Wu}). According to the National Health and Family Planning Commission, there were a total of 7,030,879 cases of notifiable infectious diseases and 19,796 deaths in 2017 (\cite{National}). Therefore, the help garnered from social media to overcome the difficulties in influenza surveillance is necessary and urgent. In this study, influenza-related tweets from Weibo are sampled to examine the capability of social media to sense a regional spectrum of symptoms. Weibo, the most popular Twitter-like service in China, has attracted 503 million users, thereby surpassing Twitter and accounting for 83.4\% of the total number of Internet users in China. Most notably, in contrast to Twitter, Weibo users are legally authenticated because of the real-name registration regulation, which further guarantees the authenticity of the samples.

It is confirmed in the present study that social media can sense a comprehensive spectrum of symptoms for influenza, especially variations across regions. Specifically, patients in southern China are more optimistic than their counterparts in the north, who, however, express more intense emotions in terms of emoticons. In the spring, patients in the south demonstrate greater odds of going to hospitals, while in the winter, the north shows more incentives for treatments. It is also revealed that the influenza duration in the south is longer than that in the north. The differences we found essentially help improve the performance of regression models in influenza surveillance. Our findings suggest that social media can be a necessary supplement to existing influenza monitoring systems.

\section{Reuslts}

To be consistent with the influenza-like illness (ILI) records and existing research, we split China geographically into two typical regions (see Fig. 1 (a)), i.e., the south and the north (\cite{Jingyang}). Purely from the ILI of both regions, the regional difference, though only in volume, can be found, as seen in Fig. 1(b). As shown in Figures. 1(c)-(d), the volume of influenza-related tweets (IRT) is positively correlated with the ILI of both the north and the south; however, the correlation is larger for the north, thereby implying the existence of differences across regions from the view of social media. In the following section, we will examine the regional differences in responses to influenza from multiple and comprehensive views. The enhancement from regional differences to regression models, which are key parts of influenza surveillance, will also be demonstrated.

\begin{figure}[h]
	\includegraphics[scale=0.45]{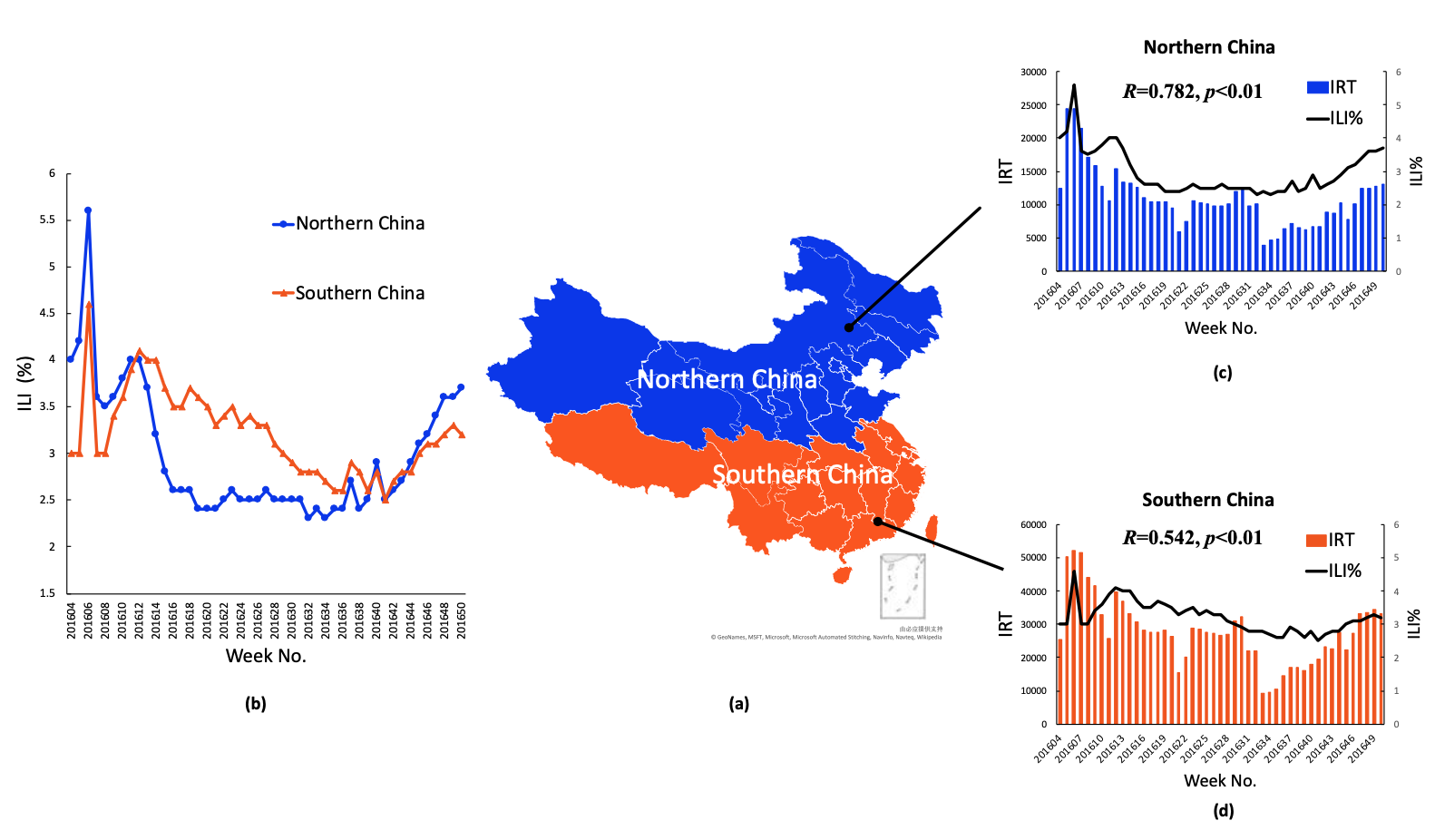}
	\caption{(a) China can be geographically split into two typical regions to be consistent with the ILI records: the south (in orange) and the north (in blue). (b) Influenza-like illness cases (ILI\%) from the Chinese Influenza Center in the south and north from the 4th to 52nd week in 2016. (c)(d) The correlation between influenza-related tweets (IRT) and ILI\% in the north and south, attaching the Pearson correlation coefficient R and asymptotic significance p-value p. The testing results are both significant, and the closer to 1 R is, the stronger the agreement.}
	\label{fig1}
\end{figure}

\paragraph{\textbf{Differences in exterior signs}} Starting from seeding terms that reflect the basic symptoms of influenza (see Methods), detailed descriptions of symptoms can be further inferred by targeting semantically similar words. In each region, by connecting seeding terms with their most similar words in the Word2vec embedding space, a semantic network can be established accordingly. As shown in Figure. 2, for each seeding term, its adjacent words represent the symptoms extracted. There were significant differences in symptoms and complications between the north and the south. By summarizing the keywords that appear separately or are cooccurring, the results show that there are more concurrent diseases in the south than in the north, and other physical discomforts excepting respiratory symptoms occur more in the south than in the north. It can also be seen from Table 1 that some common influenza symptoms appear in both the north and the south, but we can still obtain respective characteristics between different regions. For example, dengue is a disease that is mainly prevalent in the tropics and subtropics, such as Guangdong province in southern China. Our results confirm this observation because “dengue” and “myocarditis” only appear in the influenza-related words in southern China and not in those of the north. Furthermore, we found that influenza in the north is relatively intense, but patients easily recover. However, the symptoms in the south can be more serious. The abovementioned differences between regions can be well explained by environmental differences, such as in climate, temperature, and pollution, which lead to regional behaviors of the influenza virus (\cite{Lowen}\cite{Sooryanarain}). The difference in external signs revealed by social media are consistent with previous knowledge, and this information can be obtained in a real-time and low-cost manner, thereby implying that the efficiency of the existing surveillance system would be essentially enhanced.

\begin{figure}[h]
	\includegraphics[scale=0.45]{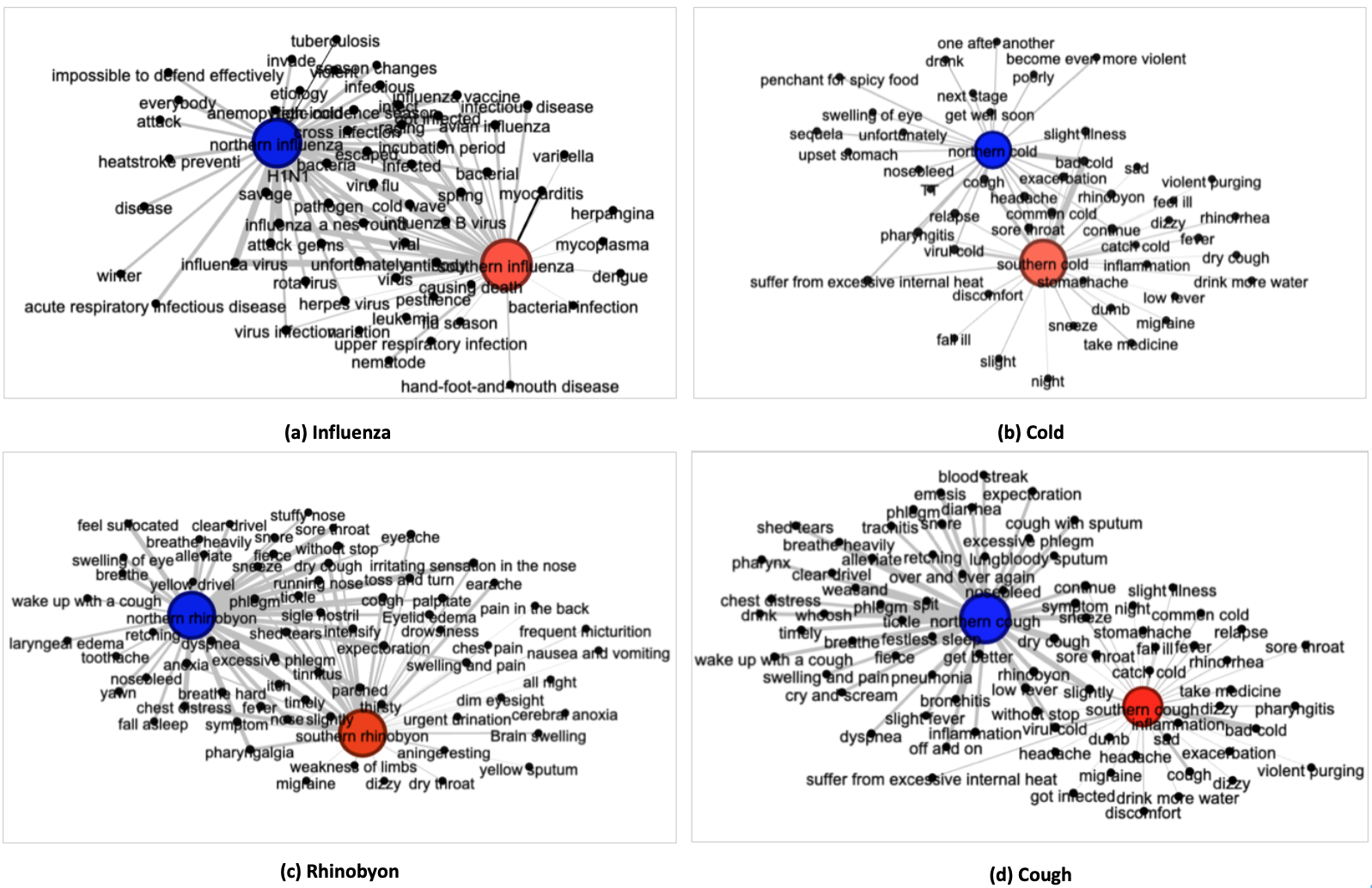}
	\caption{Word networks of typical influenza symptoms. For each region, a Word2vec model was trained independently on influenza-related tweets posted in this region, and the most similar 100 words were extracted for all the seeding words. After manually removing some ambiguous and repetitive terms, a word network was established by connecting each seeding word with its synonyms that only occur in one region. (a) Network of the seeding word ‘influenza’. (b) Network of the seeding word ‘cold’. (c) Network of the seeding word ‘rhinobyon’. (d) Network of the seeding word ‘cough’.}
	\label{fig2}
\end{figure}

\begin{table}    
	\caption{Exterior signs inferred for seeding words. Selected from Figure. 2, the most representative similar words for each seeding word are listed to better understand the differences between the south and north.}  
	\label{table1}
	\begin{center}  
		\begin{tabular}{c|m{3cm}<{\centering}|m{3cm}<{\centering}| m{3cm}<{\centering}}  
			\hline  
			 & \textbf{Northern China} & \textbf{Common} & \textbf{Southern China} \\ \hline  
			\textbf{Influenza} & winter, acute respiratory infectious disease, season changes, tuberculosis & infectious disease, spring, incubation period, virus infection, influenza vaccine, cross infection & dengue, myocarditis, hand-foot-and-mouth disease, pestilence, leukemia \\ \hline  
			\textbf{Rhinobyon} & toothache, retching, swelling around eyes, nosebleed, get better & sore throat, running nose, cough, chest distress, itch & nausea and vomiting, cerebral anoxia, dizzy, weakness of limbs, all night \\ \hline
			\textbf{Cold} & swelling around eyes, upset stomach, come in waves, become even more severe, get well soon & cough, headache, sore throat, rhinobyon, common cold & at night, stomachache, dry cough, migraine, dizzy \\ \hline
			\textbf{Cough} & intermittent cough, pneumonia, bronchitis, dyspnea, over and over again & rhinobyon, sore throat, sneeze, dry cough, inflammation & fever, stomachache, dizzy, exacerbation, migraine \\ \hline
			
		\end{tabular}  
	\end{center}  
\end{table}

\paragraph{\textbf{Differences in treatment incentives}}It is challenging with the existing influenza surveillance system to measure the incentives for treatment after getting infected because conventional data collection depends heavily on face-to-face inquiries in clinics. In social media, however, incentives for treatment can be sensed through hospital-related keywords that are frequently mentioned in tweets (see Methods). We suppose that more occurrences of hospital-related keywords inherently indicate stronger intentions for treatment. For each region, the ratio of tweets that contain hospital-related words can be accordingly defined as a value of the treatment incentive. Surprisingly, Table 2 shows that there is no significant difference across regions when neglecting the seasons ($p>0.05$). However, for the seasons of spring and winter, significant differences in treatment incentives indeed exist ($p<0.05$). Specifically, the incentives for treatment in southern China are approximately 5.0\% higher than those in the north in the spring; in winter, the incentives in the north are 10.9\% higher than those in the south. For spring and winter, the necessity of considering regional discriminations in the deployment of medical resources is thus suggested.

\begin{table}    
	\label{table2}
	\caption{Chi-square tests on treatment incentives across regions and seasons.}  
	\begin{center}  
		\begin{tabular}{m{1.5cm}<{\centering}|m{1.5cm}<{\centering}|c|c|c|c|c}  
			\hline  
			\textbf{Regions} & \textbf{Seasons} & \textbf{Spring} & \textbf{Summer} & \textbf{Autumn} & \textbf{Winter} & \textbf{Total}\\ \hline  
			\textbf{Northern China} & \textbf{Incentive (\%)} & 4.23 & 4.07 & 4.09 & 4.06 & 4.12\\ 
			\hline  
			\textbf{Southern China} & \textbf{Incentive (\%)} & 4.44 & 4.01 & 4.18 & 3.66 & 4.09\\ 
			\hline  
			\textbf{Total} & \textbf{Incentive (\%)} & 4.38 & 4.03 & 4.16 & 3.78 & 4.10\\ 
			\hline
			\multicolumn{2}{c|}{\textbf{P value}} & \textbf{0.001} & 0.366 & 0.209 & \textbf{0.000} & 0.347\\ \hline
		\end{tabular}  
	\end{center}  
\end{table}

\paragraph{\textbf{Differences in emotions}}Sophisticated signals delivered in tweets not only include descriptions of external signs of influenza but also internal feelings, e.g., emotions after becoming infected. For the clinic-based system, patients’ emotions are rarely sampled and put into records. However, from texts in social media, emotions can be comprehensively measured (\cite{Paltoglou}). To study patients’ emotional responses to influenza, we calculated the sentiment intensity of each tweet and classified them into positive and negative sentiments (see Methods). The result of the chi-square test of sentiment polarity across regions is significant ($p<0.05$), as shown in Table 3, thereby implying that regional differences in emotions do exist. In terms of negative emotions, the actual number of messages is 6\% higher than the estimated number in the north, while the opposite is true in the south. However, the regional difference in positive valence becomes trivial between the north and south. This result indicates that compared with southern China, the proportion of negative emotions is significantly higher in its northern counterpart, thereby implying that influenza infection is more likely to result in pessimism in the north.

\begin{table}    
	\caption{Chi-square tests on emotion polarity across regions.}  
	\label{table3}
	\begin{center}  
		\begin{tabular}{m{1.5cm}<{\centering}|m{1.5cm}<{\centering}|m{1.6cm}<{\centering}|m{1.5cm}<{\centering}|m{1.6cm}<{\centering}|m{1.5cm}<{\centering}}  
			\hline  
			& \multicolumn{2}{|c|}{\textbf{Positive}} & \multicolumn{2}{c|}{\textbf{Negative}} & \textbf{Total}\\ \hline  
			& \textbf{Actual Value} & \textbf{Estimated Value} & \textbf{Actual Value} & \textbf{Estimated Value} & \\ \hline  
			\textbf{Northern China} & 87264 & 87291.63 & 374173 & 371926.73 & 512836 \\ \hline  
			\textbf{Southern China} & 221142 & 221114.37 & 939864 & 942110.27 & 1299041 \\\hline  
			\textbf{Total} & 308406 & 308406 & 1314037 & 1314037 & 1811877 \\\hline
			\textbf{P value} & \multicolumn{5}{|c}{\textbf{0.000}} \\\hline		
		\end{tabular}  
	\end{center}  
\end{table}

Except for polarity in emotional expressions, the intensity of both positivity and negativity also matters. In this study, the emotional intensity or arousal can be measured with regard to the scores of terms in the emotion vocabulary (see Methods). The seasonal emotional intensity of each region can be obtained accordingly by averaging the emotional intensity of regional tweets. As seen in Figure. 3, for all seasons, the north is more intense both in the negative and positive valence, i.e., the absolute value of the average sentiment score is higher, especially in winter. This result suggests that in the north, contracting influenza leads to intense emotions, especially negative ones.

\begin{figure}[h]
	\centering
	\includegraphics[scale=0.6]{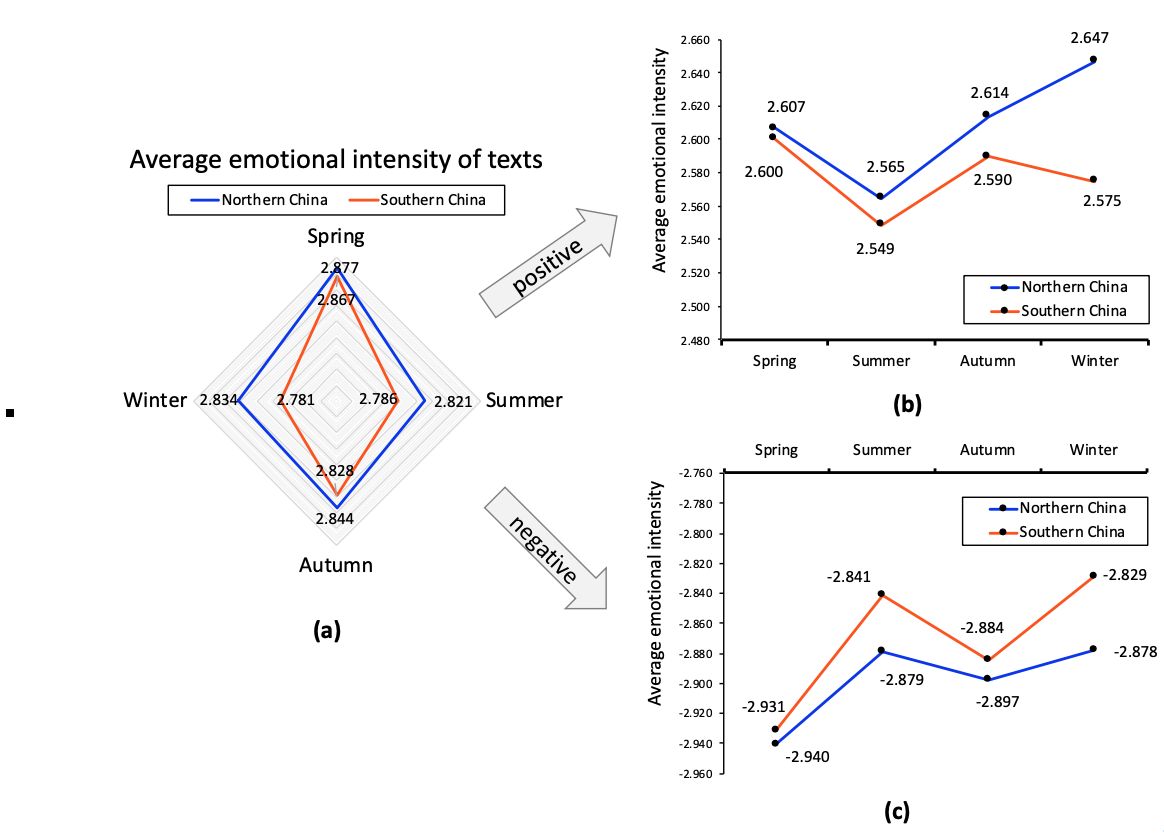}
	\caption{Emotional intensity across regions. Higher intensity implies more intensive expressions of emotions after infection. (a) Average absolute values of the emotional intensity of texts across seasons in the south and north. (b)(c) Average emotional intensities of positive and negative texts across seasons in the south and north. Unsurprisingly, the emotional intensities of positive texts are greater than 0, while the values of negative texts are less than 0.}
	\label{fig3}
\end{figure}

In contrast to sentimental terms, an emoticon is another means of emotion expression in social media (\cite{Petra}). More emoticons in tweets suggests more intensive expressions of emotions, meaning that the emotional intensity can also be valued by counting the emoticons. Herein, we manually group frequently used emoticons in Weibo (\cite{Zhao}) into four categories, in which the categories of joy and happiness represent the positive and the categories of sadness and anger represent the negative. As shown in Figure. 4, consistent with our inference from the scores of terms, negative emoticons are used more frequently than positive ones in influenza-related tweets, both in the north and south. In particular, the expressions of sadness demonstrate the highest frequency, thereby implying that negative feelings such as being sad dominate after infection. In the meantime, the frequency of joyful and angry emoticons, which essentially represent more intense emotional arousal, is significantly higher in the north than in the south. This result is also consistent with the observation regarding the view of the emotion scores of texts. The regional differences in emotional intensity indicate that the internal feelings after influenza infection can also be surprisingly impacted by region, thereby suggesting the consideration of emotion regulation or relief in medical treatments, especially for patients of the north.

\begin{figure}[h]
	\centering
	\includegraphics[scale=0.45]{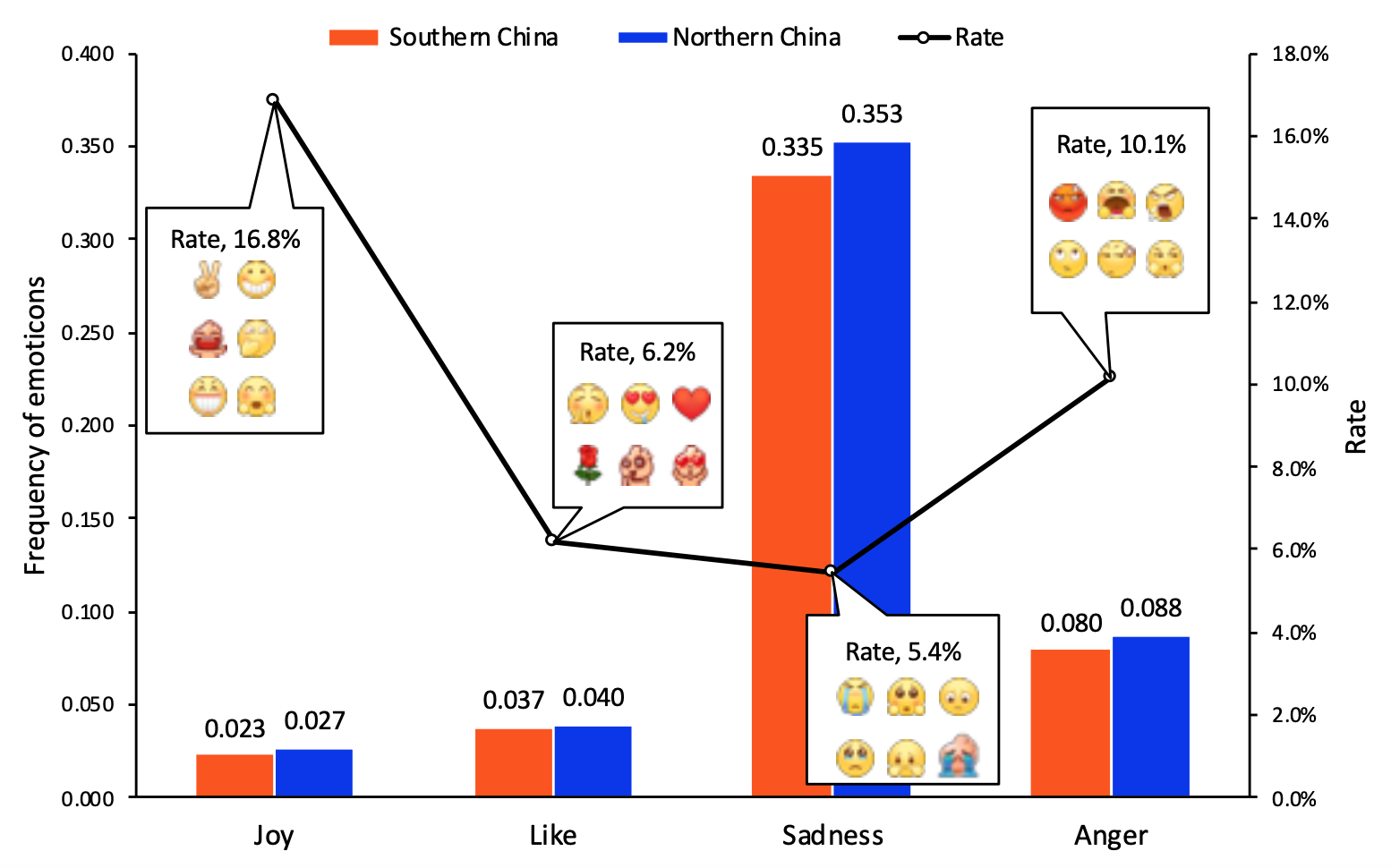}
	\caption{Emotional intensity from the examination of emoticons. Bars show the frequency of emoticons across four categories of emoticons, i.e., the average occurrence of emoticons in each tweet in the south (in orange) and the north (in blue). Rates denote the proportion of the northern emoticon frequency compared to that of the south across the four types of emoticons.}
	\label{fig4}
\end{figure}

\paragraph{\textbf{Differences in infection duration}}Distinguishing the symptoms, treatment incentives and emotions in influenza infection would intuitively lead to different durations of the disease in both regions we investigate. To confirm this directly in social media, terms reflecting the duration of influenza were extracted and sized based on their occurrence in tweets, as shown in Figs. 5(a)-(b). It is obvious that a difference in influenza duration exists between the regions, and, as we expect from the aforementioned disseminations, when compared to the north, southern China suffers a longer influenza duration. Specifically, in the word cloud map of southern China (see Figure. 5(b)), terms that indicate longer periods occur more frequently, such as “one month”, “more than one month”, and “long time”.
Because the self-healing time for influenza is basically a week (\cite{Cao}), 14 duration words (see Methods) that reflect the continuation of the flu for more than a week are manually selected to further demonstrate the differences between both regions. Tweets containing these words are named prolonged influenza-related tweets (PIRT). Figs. 5 (c)-(d) show the weekly ratio of PIRT in both the south and north; the higher the ratio, the longer the duration of influenza. It can be seen that in every season, the ratio of PIRT in southern China is generally higher than that in the north, which further confirms the significant difference in the duration even after neglecting the self-healing cases. We conjecture that because southern individuals have a longer duration of infection, they may seek medical treatments many times during influenza, while the number of influenza-related tweets posted by them may not proportionally grow. In line with this, an underestimation of real influenza occurrences might result, which can explain the relatively poorer correlation in the south (see Figure. 1 (d)).

\begin{figure}[h]
	\centering
	\includegraphics[scale=0.6]{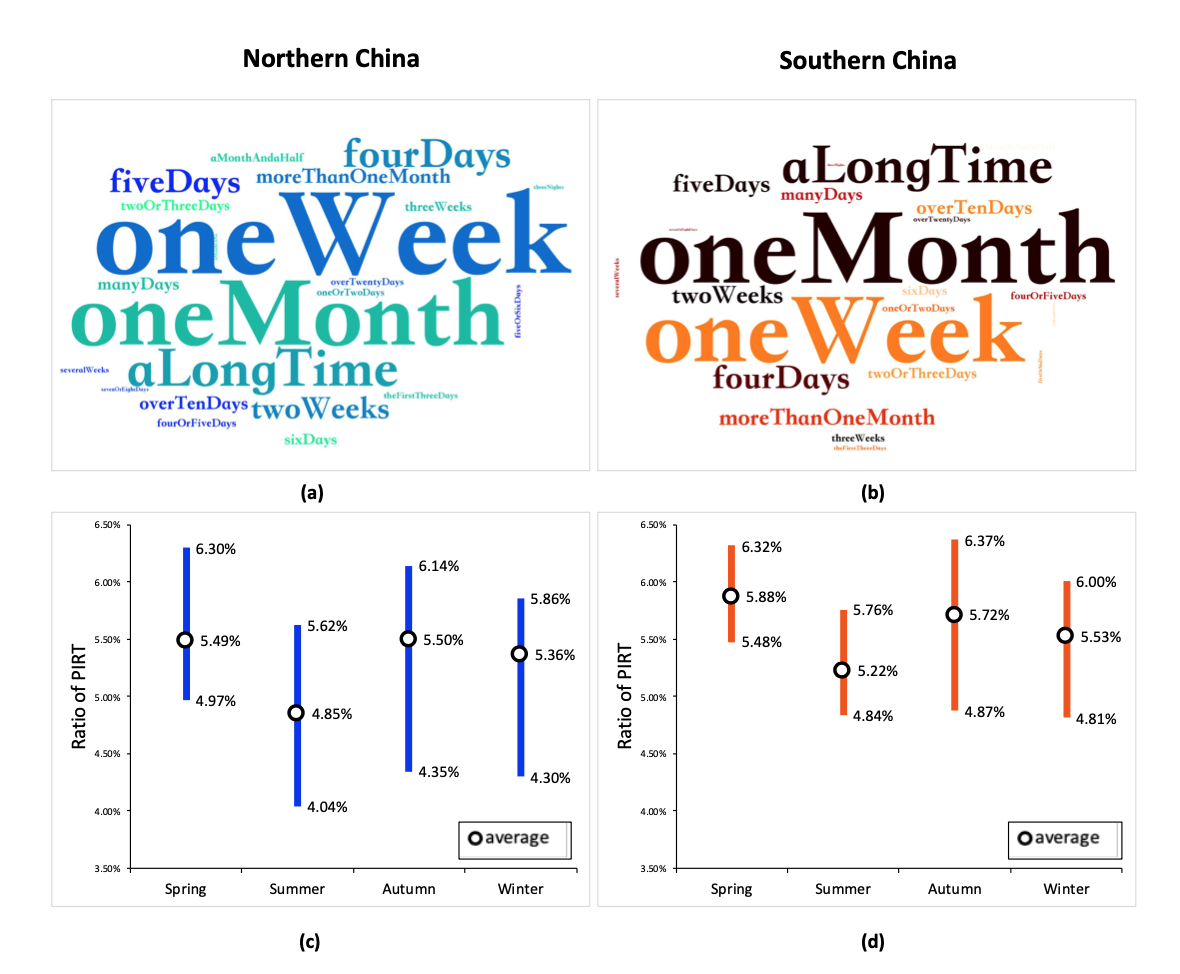}
	\caption{(a) Duration-related words in the north. Sizes of words are proportional to their occurrences in tweets posted in the north. (b) Duration-related words in the south. Similarly, sizes of words are proportional to their occurrences in southern tweets. (c) Ratios of PIRT in the north across the seasons. (d) Ratios of PIRT in the south across the seasons. }
	\label{fig5}
\end{figure}

Differences can enhance the capability of surveillance Regression models, which explain trends in the influenza rate, are key parts of the system of influenza surveillance. According to the results of various models in Table 4, influenza-related tweets in Weibo can reflect objective flu trends. In particular, models combining IRT with historical influenza data, ILI\% lagging two periods, show better results both in the north and south. Furthermore, in this paper, we first compared the interpretation of the regression model for both the north and south and found that the model performs better with northern data than with southern data. We next examined whether the difference in the duration of influenza would affect the performance of the regression model in both regions. Specifically, the explanatory power of the model is revalued after adjusting the volume of influenza-related tweets by considering the number of PIRT. That is, to adjust the IRT in week t, the PIRT in week t will be added to the IRT in week t+1, and the new IRT in week t is denoted as $Adjusted-IRT_{t}$.Table 4 shows that in the south, the explanatory power of the model has been significantly improved by the adjustment from 53.50\% to 73.10\%, but the improvement vanishes in the north. Figure. 6 shows the fitting curves before and after adjusting the IRT in the south. It can be similarly seen that there are significant improvements in the weeks where the previous fitting does not perform well (such as weeks 201608, 201614, etc.). The improvement in the regression model for the south essentially implies inspiring insights from understanding the regional differences. Indeed, with the help of social media, a more comprehensive and detailed regional spectrum of symptoms will be necessary supplements to the existing system of influenza surveillance.

\begin{table}    
	\centering
	\label{table4}
	\caption{Regression results of the baseline model, as well as before and after IRT were adjusted, for northern and southern China (see Methods). Deviance explained represents the explanatory ability of the model. RMSE is the root-mean-square deviation, and AIC is the Akaike information criterion. A higher value of deviance explained and a lower RMSE and AIC means a better model.}  
	\begin{center}  
		\begin{tabular}{m{6cm}|m{2cm}<{\centering}|c|c}  
			\hline  
			\multicolumn{4}{c}{\textbf{Northern China}} \\
			\hline  
			& \textbf{Deviance explained} & \textbf{RMSE} & \textbf{AIC}\\  
			\hline
			$ N1:log(ILI_{t}) \sim ILI_{t-2}^{***} $ & 53.2\% & 0.448 & 61.347 \\
			$ N2:log(ILI_{t}) \sim s(IRT_{t})^{***} $ & 75.0\% & 0.327 & 42.998 \\
			$ N3:log(ILI_{t}) \sim s(IRT_{t})^{**}+ILI_{t-2}^{**} $ & 82.8\% & 0.271 & 31.988 \\
			$ N4:log(ILI_{t}) \sim s(Adjusted-IRT_{t})^{**}+ILI_{t-2}^{**} $ & 82.9\% & 0.271 & 30.876 \\
			\hline
			\multicolumn{4}{c}{\textbf{Southern China}} \\
			\hline  
			& \textbf{Deviance explained} & \textbf{RMSE} & \textbf{AIC}\\  
			\hline
			$ S1:log(ILI_{t}) \sim ILI_{t-2}^{***} $ & 35.9\% & 0.367 & 43.434 \\
			$ S2:log(ILI_{t}) \sim s(IRT_{t})^{***} $ & 37.1\% & 0.363 & 43.128 \\
			$ S3:log(ILI_{t}) \sim s(IRT_{t})^{***}+ILI_{t-2}^{***} $ & 50.9\% & 0.321 & 33.372 \\
			$ S4:log(ILI_{t}) \sim s(Adjusted-IRT_{t})^{***}+ILI_{t-2}^{***} $ & 71.8\% & 0.243 & 22.313 \\
			\hline
		\end{tabular}  
	\end{center}  
\end{table}

\begin{figure}[h]
	\centering
	\includegraphics[scale=0.8]{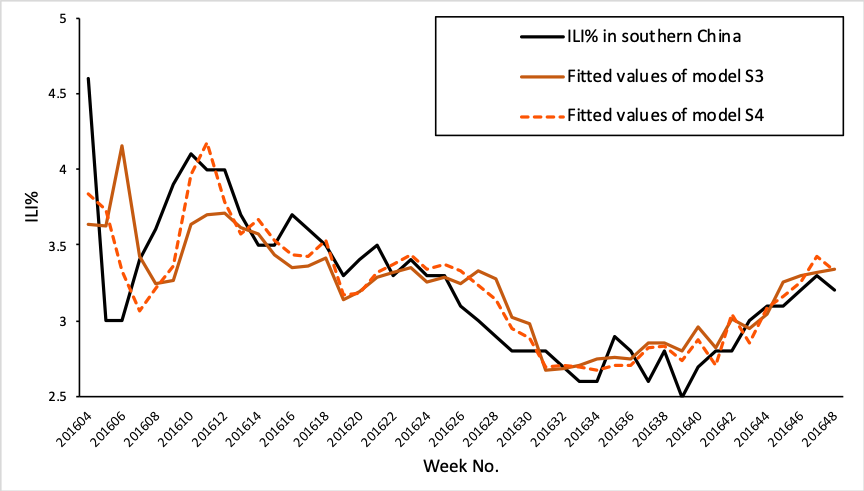}
	\caption{(a) Duration-related words in the north. Sizes of words are proportional to their occurrences in tweets posted in the north. (b) Duration-related words in the south. Similarly, sizes of words are proportional to their occurrences in southern tweets. (c) Ratios of PIRT in the north across the seasons. (d) Ratios of PIRT in the south across the seasons. }
	\label{fig6}
\end{figure}

\section{Discussion}

How to achieve effective monitoring and prevention of influenza, which is a highly contagious and rapidly mutative disease, has always been an important issue. In this paper, with the help of social media, we examine the inference of a regional spectrum of influenza symptoms, especially the differences across regions. To our knowledge, for the first time, it is confirmed that social media can sense the differences in a regional spectrum of influenza. Specifically, from perspectives such as symptom descriptions, treatment incentives, emotion expressions and illness duration, different responses in both northern and southern China are systematically and comprehensively revealed. Even more importantly, these differences can be insights into optimizing the deployment of medical resources or developing regression models in disease monitoring. Considering the inherent similarities between Weibo and other Twitter-like services, our findings can also be easily extended to other countries or circumstances.

Social media can be a new channel for the Centers for Disease Control to monitor and prevent diseases. Our results demonstrate the power of these Twitter-like services in sensing the spectrum of influenza and their surprising capabilities in disclosing regional differences in symptoms. More importantly, social media can offer insights that cannot be easily accessed through existing approaches. Interior responses, such as incentives and emotions after influenza infection, are conventionally difficult to effectively collect but offer profound clues for developing treatment plans. However, with the help of text mining and sentiment analysis, internal signals such as treatment incentives and emotional expressions can be efficiently and precisely detected in a real-time manner. From this perspective, social media can enhance the existing influenza system with its ability for comprehensive and real-time surveillance.

In particular, interesting patterns of regional differences, not only in external signs but also in internal reactions, can be deeply sensed in social media. In traditional clinic-based systems, the great expense in both cost and time makes it challenging to obtain an accurate and complete picture of influenza differences across regions. Those differences that are revealed in social media are actually helpful in monitoring and predicting influenza trends. For example, an adjusted regression model can be accordingly enhanced with a nearly 20\% improvement in performance. Our results essentially indicate that voluntary self-reports in social media can naturally and factually reflect behavioral differences among a massive number of users.

With a massive number of users being sensors anytime and anywhere, social media indeed provides important supplements to conventional influenza surveillance and treatment systems. In combination with clinical records, demographics such as age and gender would be further verified and considered in inferring the regional spectrum of symptoms in future work.

\section{Methods}

\paragraph{\textbf{Data set}}Influenza is an acute respiratory disease whose pathogen is the influenza virus. The clinical manifestations are fever, cough, headache, muscle pain, nasal congestion, sore throat, fatigue, and sometimes gastrointestinal discomfort. Therefore, our analysis in this paper is based on Sina Weibo messages containing words related to disease symptoms. The original data were obtained through an open API with Weibo, and the keywords “influenza”, “cold”, “cough”, “fever”, “sneeze”, and “rhinobyon”, which sufficiently represent the symptoms of influenza in verbal Chinese, were selected to filter the tweet stream. Finally, more than two million Weibo messages referring to influenza were obtained from 2016-1-25 to 2016-12-31. ILI (influenza-like illnesses) refer to cases with fever (body temperature higher than or equal to 38 $^{\circ}$C) and cough or sore throat, while there is a lack of other laboratory diagnostic bases. The National Influenza Center publishes reports (distributed in provinces, autonomous regions and municipalities) of influenza outbreaks from sentinel hospitals and nonnational sentinel hospitals weekly, usually with a delay of 1-2 weeks. The influenza surveillance network in China is divided into two monitoring areas, namely, north and south, so we can easily obtain the data for northern China and southern China accordingly. The northern and southern ILI datasets (ILI\%) used in this paper were provided by the Chinese Influenza Center (http://www.chinaivdc.cn) and are from the 4th week to the 52nd week of 2016, which is a sampling period that is consistent with the tweets from Weibo. All the data in this study can be freely downloaded through https://doi.org/10.6084/m9.figshare.7545203.v1. 

\paragraph{\textbf{Identify influenza-related tweets}}We achieved the goal of our research by creating a linear support-vector machaine (SVM), an established model of data in machine learning (\cite{Cortes}), which works well on binary classification problems. Specifically, in the context of this study, tweets containing influenza-related keywords were further refined by classifying them into either truly describing influenza or noise. To initiate the classification task, a manually labeled corpus of 4,210 tweets, which will be divided into a training set and testing set later, was first established by well-trained coders. In the labeled corpus, influenza-related tweets were labeled as positive and the others were identified as negative. Unlike text classification in English, word segmentation is the first step in Chinese text processing. In this paper, jieba, which is a python library, was used to perform the word segmentation. The result of the word segmentation includes all terms in the training set, which has a very high dimensionality (approximately 10 thousand dimensions) with many possibly irrelevant features for the classification. To reduce dimensions, we tried the information gain (IG) method to calculate the score of every term and sorted them by the IG score in reverse order. The accuracy of classification under different dimensions was calculated both in the training and testing sets, and the dimension with the highest accuracy was chosen. After that, TF-IDF (term frequency-inverse document frequency) was used to convert the text into a vector (\cite{Forman}), which was the final input of the SVM classification.

We used the libSVM package to train our classifier and chose the RBF core as the kernel function. RBF can map nonlinear samples to high-dimensional space, which is more conducive to solving nonlinear problems. The final evaluation of the SVM classifier shows 80.9\% accuracy, 79.9\% precision and 78.0\% recall. We used the classifier to label all Weibo messages we mined in 2016, and the data labeled as positive were selected for the next analyses as IRT.

\paragraph{\textbf{Value treatment incentives}}We used hospital-related keywords in Table 5 to filter a selection of tweets that were supposed to reflect intentions for treatments. Then, we calculated the proportion of the number of tweets containing those keywords in the total number of tweets in the north and south; tweets can also be grouped by seasons such as spring and winter. The proportion results are considered as the treatment incentives in each season.

\begin{table}    
	\caption{Words denoting treatment incentives}  
	\label{table5}
	\begin{center}  
		\begin{tabular}{p{12cm}}  
			\hline  
		    \multicolumn{1}{c}{\textbf{Hospital-related keywords}}\\ 
		    \hline
		    Hospital, outpatient service, emergency treatment, draw blood, blood test, register, test, transfusion, prescribe, chest X-ray, doctor, in hospital, take an injection, pick up the medicine, return visit examination, emergency clinic, queue, make a definite diagnosis, urine test, in treatment, pneumonia, pay the fees\\ \hline   
		\end{tabular}  
	\end{center}  
\end{table}

\paragraph{\textbf{Value the emotional intensity of texts}}We chose a dictionary-based method to quantify the emotional value of the text because sentiment intension is necessary. Shen's dictionary-based method of inferring microblogs’ emotional orientations is chosen in this study (\cite{Yang}). This method is based on emotional words, degree adverbs, and the relative position of words in the tweets for emotional scoring. We used Shen’s method to calculate the emotional score of each Weibo message. To make the results follow a normal distribution, we only considered messages whose scores were between -100 and 100 and removed those of exceptional values. A tweet was considered to be positive if the $score > 0$, negative if the $score<0$, neutral if the score = 0. Neutral messages were not included within our analysis.

Select duration-related words We manually selected several seeding words first and then used the Word2vec embedding model to obtain semantically similar words to those seeding words. After further manual screening, we selected 14 words denoting longer times, which can be found in Table 6.

\begin{table}    
	\label{table6}
	\caption{Words denoting longer time.}  
	\begin{center}  
		\begin{tabular}{p{12cm}}  
			\hline  
			\multicolumn{1}{c}{\textbf{Time words}}\\ \hline
			Over ten days, two weeks, three weeks, half a month, over twenty days, many days, several weeks, a month and a half, one month, more than one month, two months, several months, a month or so, a long time\\ \hline   
		\end{tabular}  
	\end{center}  
\end{table}

\paragraph{\textbf{Generalized Additive Model (GAM) }}In statistics, a generalized additive model (GAM) is an extended linear model in which the linear predictor depends linearly on the unknown smooth functions of some predictor variables, and its interest therefore focuses on the inferences of these smooth functions (\cite{Marra}). Herein, we considered the model with only influenza historical data as the baseline solution and built a series of variant upgradations by progressively adding variables. Specifically, the GAM employed in the present study can be formally defined as follows:

 $ ILI_{t} \sim Negative Binomial (\mu_{t},\kappa) $
 
 $log(\mu_{t})=\beta_{0}+\beta_{1}s(IRT_{t})+\beta_{2}\mu_{t-2}+\epsilon_{t}$,

where $\mu_{t}$ denotes the ILI value for the period t; $\mu_{t-2}$ denotes the ILI value two weeks before period t, which is the latest influenza history data we can access from the surveillance network; $s(IRT_{t})$ is the smooth spline of the influenza-related tweets; and $\epsilon_{t}$ is the residual term for the model. For the evaluation of the model, the deviance explained by the model represents its explanatory ability in influenza surveillance. RMSE is the root-mean-square error, which is a frequently used measure of the differences between estimations (sample or population values) predicted by a model or an estimator and the real values observed. In general, a lower RMSE is better than a higher RMSE. AIC stands for the Akaike information criterion, which is an estimator of the relative quality of statistical models for a given set of data. The AIC estimates the relative information loss by a given model, and the less information a model loses, the higher the performance of that model.

\section*{Acknowledgement}
This work was supported by the National Natural Science Foundation of China (Nos. 71771010, 71471008 and 71871006).


\begin{thebibliography}{1}
	\bibitem{Florian} Florian, K., Gavin, J. D. S., Ron, A. M. F., Malik P., Katherine, K., et al. Influenza. Nature Reviews Disease Primers 4, 1-21(2018).
	\bibitem{Cameron} Cameron, Z., Kristopher, M. F., Oliver, M. C., Nathan, H., Mahendra, P., Mikhail, P. . Urbanization affects peak timing, prevalence, and bimodality of influenza pandemics in Australia: Results of a census-calibrated model. Science Advances 4, eaau5294 (2018).
	\bibitem{Nguyen} Nguyen-Van-Tam, J. S., Hampson, A. W. . The epidemiology and clinical impact of pandemic influenza. Vaccine 21, 1762-1768 (2003).
	\bibitem{Beaut} Beauté, J., Zucs, P., Korsun, N., et al. Age-specific differences in influenza virus type and subtype distribution in the 2012/2013 season in 12 European countries. Epidemiology \& Infection 143, 2950-2958 (2015).
	\bibitem{Rolfes} Rolfes, M. A, Foppa, I. M., Garg, S., Flannery, B., Brammer, L., et al. Annual estimates of the burden of seasonal influenza in the United States: A tool for strengthening influenza surveillance and preparedness. Influenza and Other Respiratory Viruses 12, 132-137 (2018).  
	\bibitem{Bednarska} Bednarska, K., Hallmann-Szelinska, E., Kondratiuk, K., Brydak, L. B. . Influenza surveillance. Postepy Higieny I Medycyny Doswiadczalnej 70, 313-318 (2016).
	\bibitem{Ginsberg} Ginsberg, J., Mohebbi, M., Patel, R., Brammer, L., Smolin-ski, M. \& Brilliant, L. . Detecting influenza epidemics using search engine query data. Nature 457, 1012-4 (2009).
	\bibitem{Lazer} Lazer, D., Kennedy, R., King, G. \& Vespignani, A. . The parable of google flu: traps in big data analysis. Science 343, 1203-5 (2014).
	\bibitem{Abbasi} Abbasi, \& Ali, M. . Understanding social media users via attributes and links. ACM SIGWEB Newsletter Summer 2015, 1-1 (2015). 
	\bibitem{Charles} Charles-Smith, L. E., Reynolds, T. L., Cameron, M. A, Conway, M, Lau, E. H. Y. \& Olsen, J. M. . Using social media for actionable disease surveillance and outbreak management: a systematic literature review. Plos One 10, e0139701 (2015). 
	\bibitem{Moritz} Moritz, W. , Vasileios, L. , Ingemar, J. C. \& Richard, P. . The added value of online user-generated content in traditional methods for influenza surveillance. Scientific Reports 8, 1-9 (2018).
	\bibitem{Fred} Fred, S. L., Mohammad, W. H., Cesar, L. C., Matthew, B. \& Mauricio, S. Improved state-level influenza nowcasting in the United States leveraging Internet-based data and network approaches. Nature Communications 10, 1-10 (2019). 
	\bibitem{Marque} Marque-stoledo, C. D. A., Degener, C. M., Vinhal, L., Coelho, G., Meira, W. \& Codeço, C.T., et al. Dengue prediction by the web: tweets are a useful tool for estimating and forecasting dengue at country and city level. Plos Neglected Tropical Diseases 11, e0005729 (2017).
	\bibitem{Veruska} Veruska, S. . The role of positive emotion and contributions of positive psychology in depression treatment: systematic review. Clinical Practice \& Epidemiology in Mental Health 9, 221-237 (2013).
	\bibitem{Askelund} Askelund,A. D., Schweizer, S.,  Goodyer I. M., Harmelen, A. L. V. . Positive memory specificity is associated with reduced vulnerability to depression. Nature Human Behaviour 3, 48-56 (2019). 
	\bibitem{Mellblom} Mellblom, A. V., Ruud, E., Loge, J. H. \& Lie, H. C. . Do negative emotions expressed during follow-up consultations with adolescent survivors of childhood cancer reflect late effects?. Patient Education and Counseling 100, 2098-2101 (2017).
	\bibitem{Volkova} Volkova, S., Charles, L. E., Harrison, J. \& Corley, C. D. . Uncovering the relationships between military community health and affects expressed in social media. Epj Data Science 6, 9(2017).
	\bibitem{Yu} Yu, B. H., Alonso, W. J., Feng, L., Tan, Y., Shu, Y. \& Yang, W. . Characterization of regional influenza seasonality patterns in china and implications for vaccination strategies: spatio-temporal modeling of surveillance data. Plos Medicine 10, e1001552 (2013).
	\bibitem{Wu} Wu, S., Wei, Z., Greene, C. M., Yang, P., Su, J. \& Song, Y., et al. Mortality burden from seasonal influenza and 2009 h1n1 pandemic influenza in beijing, china, 2007-2013. Influenza and Other Respiratory Viruses 12, 88-97 (2018).
	\bibitem{National} National Health and Family Planning Commission of the People 's Republic of China. 2017 National Statutory Infectious Diseases. http://www.nhfpc.gov.cn/jkj/s3578/201802/de926bdb046749abb7b0a8e23d929104.shtml.
	\bibitem{Jingyang} Jingyang, Z., Hua, Y., Hengjian, C., et al. Geographic Divisions and Modeling of Virological Data on Seasonal Influenza in the Chinese Mainland during the 2006–2009 Monitoring Years. Plos One 8, e58434(2013).
	\bibitem{Lowen} Lowen, A. C., Mubareka, S., Steel, J., et al. Influenza Virus Transmission Is Dependent on Relative Humidity and Temperature. Plos Pathogens 3, 1470-1476 (2007). 
	\bibitem{Sooryanarain} Sooryanarain, H., Elankumaran, S. . Environmental Role in Influenza Virus Outbreaks. Annual Review of Animal Biosciences 3, 347-373 (2015).
	\bibitem{Paltoglou} Paltoglou, G., Thelwall, M. . Twitter, MySpace, Digg: Unsupervised Sentiment Analysis in Social Media. ACM Transactions on Intelligent Systems \& Technology 3, 1-19 (2012).
	\bibitem{Petra} Petra, K. N., Jasmina, S., Borut, S., Igor, M., \& Matjaz, P. . Sentiment of emojis. Plos One 10, e0144296 (2015).
	\bibitem{Zhao} Zhao, J., Dong, L., Wu, J., Xu, K. . Moodlens: An emoticon-based sentiment analysis system for Chinese tweets. In: Proc. 18th ACM SIGKDD Intl. Conf. on Knowledge Discovery and Data Mining. ACM, 1528-1531 (2012). 
	\bibitem{Cao} Cao, B., Li, X. W., Mao, Y., et al. Clinical Features of the Initial Cases of 2009 Pandemic Influenza A (H1N1) Virus Infection in China. New England Journal of Medicine 361, 2507-2517 (2009).
	\bibitem{Cortes} Cortes, C. \& Vapnik, V. . Support-vector networks. Machine learning 20, 273–297 (1995). 
	\bibitem{Forman} Forman, G. . BNS feature scaling:an improved representation over tf-idf for svm text classification. ACM Conference on Information.
	\bibitem{Yang} Yang, S., Shuchen, L., Ling, Z., et al. Emotion mining research on micro-blog. IEEE Symposium on Web Society (2009).
	\bibitem{Marra} Marra, G. \& Wood, S. N. . Coverage properties of confidence intervals for generalized additive model components. Scandinavian Journal of Statistics 39, 53-74 (2012).
	
\end{thebibliography}
\end{document}